# Sur la contamination de la lumière des étoiles par de la lumière diffusée à très faible angle.

Frédéric Zagury (fzagury@wanadoo.fr)
02210 Saint Rémy Blanzy, France

## Résumé:

La lumière diffusée qui contamine le spectre des étoiles rougies à une dépendance en $1/\lambda^n$ ($n \geq 4$), caractéristique de particules petites devant la longueur d'onde. Il est surprenant que ces particules, qui doivent diffuser de façon quasi-isotrope, ne contribuent pas à la brillance des nébuleuses. L'importance de la lumière diffusée par ces petites particules dans le spectre d'une étoile rougie s'explique seulement si la diffusion est cohérente. L'intensité diffusée est alors proportionnelle au carré du nombre de diffuseurs. Vue par l'observateur, elle est concentrée dans un angle inférieur à $10^{-8}"$ de l'étoile. Les diffuseurs peuvent être des petits grains, ou, des atomes ou molécules du gaz. L'hydrogène satisfait remarquablement aux conditions que les observations imposent aux diffuseurs.

## Abstract:

The scattered starlight which contaminates the spectrum of a reddened star depends on $\lambda$ as $1/\lambda^n$ ($n \geq 4$), and must be attributed to small particles. Surprisingly these particles, which have a quasi-isotropic phase function, do not participate to the brightness of the nebulae. The importance of the intensity of the scattered light present in the spectrum of reddened stars can be explained only if the scattering is coherent. The intensity of the scattered light is then proportional to the square of the number of particles, and is concentrated within an angle of $10^{-7}"$ from the star. The scatterers can be small grains, or, atoms or molecules from the gas. Hydrogen satisfies all conditions imposed to the scatterers.

## 1. Introduction

J'ai repris dans une note précédente [1] différents arguments qui montrent que la conception standard de l'extinction interstellaire, sur laquelle repose tous les modèles de poussière interstellaire, ne résiste pas à la confrontation avec certaines observations. Si cette théorie est fausse, il faut, pour expliquer le spectre UV des étoiles rougies, que la lumière qui nous parvient de ces étoiles soit contaminée dans une proportion non négligeable par de la lumière diffusée.

Nous savons déjà [1] que cette diffusion est restreinte à un angle faible, qui doit être petit devant la seconde d'arc. L'objet de cette note est de mieux caractériser la lumière diffusée, sa nature et celle des diffuseurs. Cette caractérisation annonce des directions nouvelles de recherche dont il est question dans la section 6. Je résume les principaux apports de cette étude dans la conclusion.

## 2. Le spectre de la lumière diffusée

L'importance relative de la lumière diffusée qui contamine le spectre d'une étoile rougie peut être estimé à partir du spectre UV d'étoiles de rougissement intermédiaire (E(B-V) de l'ordre de 0.1-0.4) pour lesquelles la lumière diffusée est surtout présente dans l'UV lointain tandis que la lumière directe de l'étoile reste la composante principale du spectre dans le proche UV, jusque dans la région du bump. Pour ces étoiles la lumière diffusée peut représenter jusque 15-20% de la lumière directe et non-éteinte de l'étoile [2]. A titre de comparaison, la lumière que nous recevons directement de l'étoile à la longueur d'onde $\lambda$ est éteinte de environ $e^{-2E(B-V)/\lambda}$ ($\lambda$ en $\mu$m) si l'extinction linéaire du visible s'étend à l'UV; pour $1/\lambda \sim 6\mu m^{-1}$ et E(B-V)=0.5 la lumière directe est de l'ordre de 0.1% du flux de l'étoile corrigé du rougissement. Ainsi, au-delà du bump, la lumière diffusée domine largement le spectre observé de l'étoile.

La diffusion s'effectuant à très faible angle, il est clair que lumière diffusée et lumière directe traversent les mêmes milieux, et sont donc éteints de façon semblable. En corrigeant le spectre d'une étoile rougie, normalisé par celui d'une étoile non rougie de même type spectral (ce que j'appelle le "spectre réduit" de l'étoile dans la suite), de l'extinction, on doit donc obtenir une constante (1 si étoile rougie et étoile non rougie sont identiques) sur la partie visible du spectre, ou la lumière directe reste dominante, et un spectre caractéristique du processus de diffusion dans l'UV lointain.

Il est nécessaire, pour séparer au mieux lumières directe et diffusée, de disposer de spectres complets d'étoiles du proche infra-rouge à l'UV lointain. J'ai récupéré du catalogue établi par J.F. Le Borgne (Observatoire de Nice), le spectre visible de l'étoile rougie HD46223 (E(B-V)~0.5) et celui de l'étoile faiblement rougie de même type spectral HD269698 (E(B-V)~0.1). Ces deux étoiles ont aussi été observées dans l'UV par le satellite IUE (International Ultraviolet Explorer). On peut donc établir le spectre réduit de HD46223 du proche infra-rouge jusqu'à l'UV lointain.

Si on laisse de côté la région du bump, le spectre réduit de HD46223 multiplié par $e^{2E(B-V)/\lambda}$ pour le corriger de l'extinction, est une fonction linéaire de $1/\lambda^4$ dans l'UV lointain [3]. En d'autres termes, la lumière diffusée croit comme $1/\lambda^4$.

Le spectre réduit de HD46223, en dehors de la région du bump, est bien reproduit (figure 1) par ses deux composantes: l'extinction de la lumière directe (proportionnel à $e^{-2E(B-V)/\lambda}$) et la composante de lumière diffusée (proportionnel à $\lambda^{-4}e^{-2E(B-V)/\lambda}$).

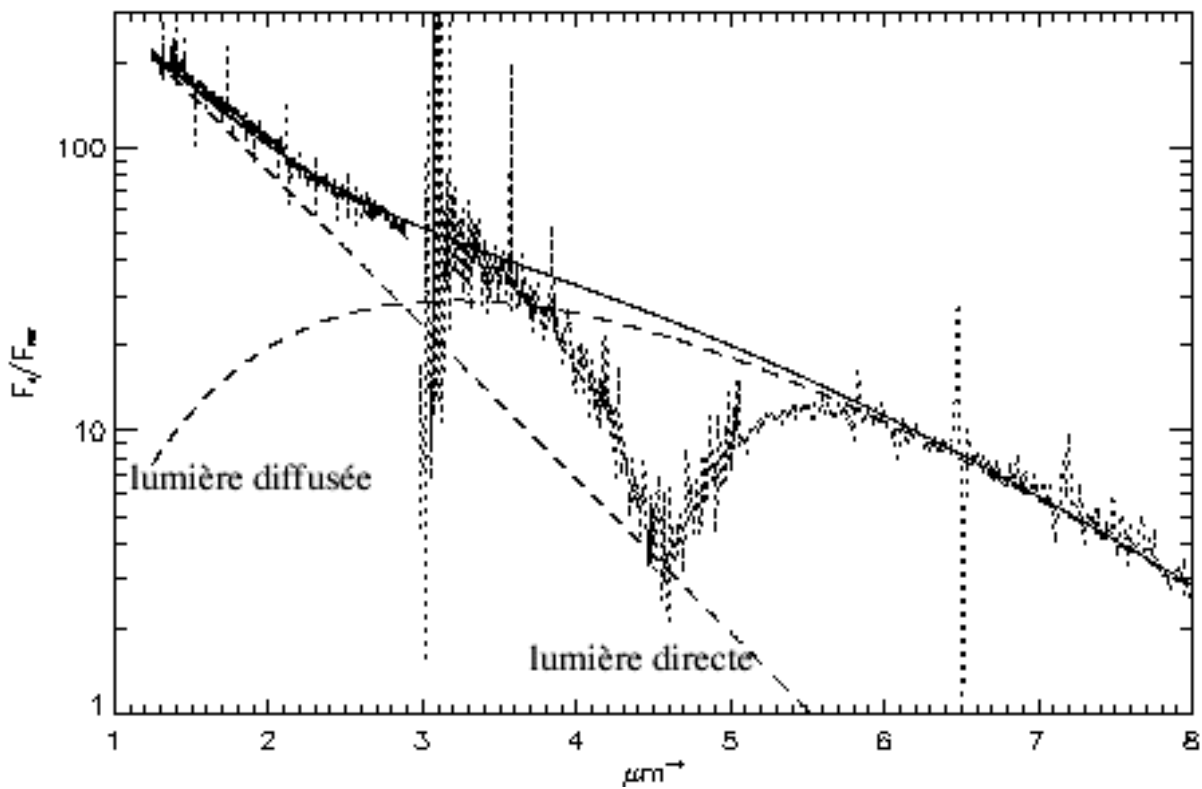

Figure 1 : spectre réduit de HD46223 et sa décomposition (hors région du bump) en lumière directe venant de l'étoile et lumière diffusée. La courbe en trait plein est la somme des deux et reproduit bien le spectre.

## 3. Relation entre les parties visibles et UV lointain de la courbe d'extinction

Si la lumière diffusée dépend de la longueur d'onde en $\lambda^{-4}$, le produit de $\lambda^4$ et du spectre réduit d'une étoile fortement rougie sera, dans l'UV lointain ou la lumière diffusée domine la lumière directe, proportionnel à

$e^{2E(B-V)/\lambda}$.

Pour un certain nombre d'étoiles j'ai comparé [4] la valeur $E_{uv}$ de E(B-V) obtenue en ajustant la partie UV lointaine du produit (spectre réduit)x$\lambda^4$ à une exponentielle $e^{-2E_{uv}/\lambda}$, à celle, $E_{vis}$, obtenue à partir du type spectral de l'étoile et de ses magnitudes visibles. L'accord entre les deux valeurs est excellent, avec une tendance pour $E_{uv}$ à être supérieur à $E_{vis}$. Cette tendance s'explique puisque la présence de lumière diffusée dans le spectre visible d'une étoile change légèrement la pente de l'extinction linéaire dans le sens d'une valeur observée de E(B-V) inférieure à sa valeur réelle [3, 4].

Cette relation très générale que l'on trouve entre les parties visible et UV lointain de la courbe d'extinction est totalement incompatible avec la théorie standard, pour laquelle extinctions visible et UV sont produites par différents types de grains, et varient de façon indépendante, sans logique apparente [5].

## 4. Diffusion à angle nulle et diffusion par les nébuleuses

Dans l'UV, la loi de diffusion par une nébuleuse, avec un angle de diffusion non-nul, est linéaire en $\lambda^{-1}$ [1]. Les particules qui sont responsables de la brillance UV d'une nébuleuse ne sont donc pas les mêmes que celles responsables de la diffusion à très faible angle associée au spectre d'une étoile rouge, qui diffusent en $\lambda^{-4}$.

Une diffusion en $\lambda^{-1}$ ne peut être produite que par des grains gros devant la longueur d'onde, qui diffusent la lumière vers l'avant. Une diffusion en $\lambda^{-4}$ (diffusion Rayleigh [6]) est caractéristique de particules petites devant la longueur d'onde, qui diffusent de façon quasi-isotrope. Il faut donc deux types de particules pour expliquer le spectre des nébuleuses et la contamination de la lumière d'une étoile par de la lumière diffusée. Cela conduit à des objections de taille :
    -comment une diffusion isotrope peut-elle être importante vers l'avant et négligeable à angle de diffusion plus large?
    -comment cette diffusion peut-elle être plus importante, vers l'avant, que la diffusion par de gros grains qui eux diffusent vers l'avant et sont responsables de la brillance dans l'UV des nébuleuses?

## 5. Diffusion cohérente de la lumière à faible angle de diffusion

Je n'ai trouvé qu'une réponse aux questions précédentes. Si la lumière diffusée est en première approximation proportionnelle au nombre de diffuseurs, elle devient proportionnelle au carré de ce nombre si la diffusion se fait en phase. En d'autres termes, si la différence de marche entre 2 trajets possibles de la lumière diffusée est petite devant la longueur d'onde, et si les diffuseurs sont identiques, le nombre de diffuseurs intervient, dans l'intensité diffusée reçue par un observateur, par son carré.

L'étoile peut être considérée comme un ensemble de sources ponctuelles; chacune de ces sources voit sa lumière diffusée en phase vers l'observateur par les particules suffisement proches de l'axe source-observateur, ce, quelle que soit la répartition des particules. L'intensité reçue est proportionnelle au carré, $N^2$, du nombre de particules. L'intensité totale reçue, en supposant N constant d'une source à l'autre, reste proportionnelle à $N^2$. La condition cherchée est donc que les différences de longueur entre la distance observateur-étoile et les trajets de la lumière diffusée soient petites devant la longueur d'onde.

Soit $l_0$ la distance nuage-observateur, $d_0$ la distance étoile-nuage dans la direction de l'observateur, D la distance étoile-observateur (D=$l_0$+$d_0$), $\theta$ la distance angulaire vue de l'observateur à l'intérieur de laquelle la lumière diffusée est cohérente. La condition de cohérence s'exprime par:

$$\theta << 5.10^{-8}" \left(\frac{\lambda}{2000\text{Å}}\right)^{0.5} \left(\frac{100pc}{l_0}\right)^{0.5} \left(\frac{d_0}{D}\right)^{0.5} \qquad (1)$$

$10^{-8}"$ est donc un ordre de grandeur typique de la distance angulaire à l'étoile vue par l'observateur et à l'intérieur de laquelle la lumière diffusée est cohérente. Pour un nuage situé à 100 pc, cela représente une surface de $10^{14}$ cm$^2$ (soit 100 km de rayon) environ. Si $\beta$ est la proportion de diffuseurs relativement au nombre d'atomes d'hydrogène, pour une colonne densité de $10^{20}$ H/cm$^2$, l'intensité diffusée sera $\beta 10^{34}$ fois ce qu'elle serait si la diffusion était incohérente.

# 6. Conséquences et prospectives

## 6.1 Nature des particules responsables de la diffusion

Il a été suggéré par T. Lehner, de l'Observatoire de Meudon, que le gaz pouvait être responsable de cette diffusion. L'idée est séduisante, en particulier parce que le gaz est l'élément le plus important du milieu interstellaire et parce que cela garantit que les diffuseurs sont identiques. Suivant Rayleigh [6], un ensemble d'atomes ou de molécules devrait diffuser proportionnellement à $\lambda^{-4}e^{-b\lambda^{-4}}$ ($b$ est une constante positive). Pour que cela soit consistant avec les observations il faut a priori que ($b\lambda^{-4}$) soit petit dans l'intervalle de longueur d'onde considéré. Si la diffusion est faite par l'hydrogène, la symétrie de l'atome (ou de la molécule) pourrait introduire un terme en $1/\lambda^6$ (en plus du $1/\lambda^4$) dans l'expression de la section efficace.

D'après Sellgren [7], la présence de petits grains est nécessaire pour expliquer le spectre infra-rouge des nuages interstellaires. Ces grains, s'ils existent, peuvent aussi être responsables de la diffusion cohérente.

## 6.2 Colonne densité de diffuseurs

On peut estimer les colonnes densités de gaz (ici dans le cas de l'atome d'hydrogène) ou de poussières nécessaires pour justifier l'amplitude de la diffusion cohérente.

Soit $\sigma_0 = \sigma \left( \dfrac{\lambda}{2000\text{Å}} \right)^4$, ou $\sigma$ est la section efficace des particules exprimée en cm$^2$.

Si il y a N particules par cm$^2$ entre l'étoile et l'observateur et que la surface de diffusion est S, le rapport entre lumière diffusée reçue par cm$^2$ et flux de l'étoile est de l'ordre de (équation (7) de [9]) :

$$\frac{\sigma(NS)^2}{4\pi d_0^{\,2}} \left( \frac{D}{d_0} \right)^2 \approx \sigma N^2 \theta^4 l_0^{\,2} \left( \frac{D}{d_0} \right)^2 \qquad\qquad (2)$$

Un rapport de 15% (section 2) entre l'intensité diffusée par les particules et le flux de l'étoile corrigé du rougissement, est obtenu pour:

$$N \approx 4.10^{-10}\text{cm}^{-2} \left( \frac{1''}{\theta} \right)^2 \left( \frac{d_0}{D} \right) \left( \frac{100\text{pc}}{l_0} \right) \left( \frac{\lambda}{2000\text{Å}} \right)^2 \sigma_0^{\,-0.5} \qquad (3)$$

En utilisant (1), et en prenant $\lambda \approx 1500\text{Å}$ [2], les diffuseurs doivent satisfaire:

$$N \gg 10^7 \sigma_0^{\,-0.5} \text{cm}^{-2} \qquad\qquad (4)$$

Pour l'hydrogène $\sigma_0 = 2.310^{-28}$ se calcule à partir de la polarisabilité atomique. Pour un grain de poussière sphérique, $\sigma_0 = 3.210^{-8} \left( \dfrac{a}{100\text{nm}} \right)^6$ dépend de la taille (a) du grain et est donné par Van de Hulst [8].

Ainsi on trouve respectivement pour l'hydrogène et pour des grains de petite taille les conditions :

$$N_H \gg 610^{20}\text{cm}^{-2} \quad (5)$$

$$N_{grain} \gg 10^7 \left( \frac{100\text{nm}}{a} \right)^3 \text{cm}^{-2} \qquad\qquad (6)$$

L'ordre de grandeur de colonne densité donné par (5) correspond bien aux valeurs trouvées dans le milieu interstellaire. Il est aussi remarquable que la colonne densité limite de (5), $N_H = 6\ 10^{20}$ cm$^{-2}$ (soit E(B-V) de l'ordre de 0. 1), corresponde au seuil observé (voir [1]) au-delà duquel le composante de diffusion est détectée. L'hypothèse selon laquelle l'hydrogène serait l'agent de la diffusion cohérente, suggérée par Thierry Lehner (Observatoire de Meudon), apparait donc tout à fait plausible.

## 6.3 Ajustement analytique des courbes d'extinction

*« Donnez-moi quatre paramètres et je vous dessine un éléphant; donnez m'en un cinquième et il aura une trompe.»*
*citation attribuée à Joseph Bertrand, mathématicien français (1822-1900)*

L'ajustement analytique des courbes d'extinction UV se fait traditionnellement selon la décomposition de Fitzpatrick et Massa [10] qui engage 5 à 6 paramètres (dont 3 pour la région du bump). Cette décomposition, de nature purement mathématique, n'a aucune valeur physique, sauf peut-être pour la région du bump si il s'agit bien d'une absorption par un type de particules particulier. La validité de la décomposition proposée par Fitzpatrick et Massa est strictement restreinte au domaine UV: j'ai pu constater que dans certains cas le prolongement dans le visible de la paramétrisation diverge de la réalité observée. Quelle valeur prédictive peut-on accorder à une paramétrisation dont la pertinence ne dépasse pas les bornes du domaine UV pour lequel elle a été conçue? Faut-il s'étonner qu'aucune relation n'ai été trouvée entre les paramètres des différentes parties de la courbe d'extinction, entre le visible et l'UV lointain en particulier?

La séparation entre lumière diffusée et lumière directe, la relation qui existe entre extinction visible et extinction UV-lointain, montrent qu'il est possible de donner un sens physique à chacune des parties de la courbe d'extinction et doit permettre de réduire le nombre des paramètres nécessaires à ajuster les courbes d'extinction [3]. Il me semble possible qu'une étude systématique des variations de la courbe d'extinction, étendue du visible à l'UV, d'un grand nombre d'étoiles derrière un même nuage conduise de mieux comprendre les détails de la courbe d'extinction et d'en fixer le nombre exact de degrés de liberté.

## 6.4 Le bump à 2200 Å

Les sections précédentes permettent de comprendre, par une analyse croisée de différentes observations, la partie UV lointaine de la courbe d'extinction. Bien que l'interprétation traditionnelle du bump pourrait s'appliquer à ce nouveau contexte, j'ai eu l'occasion de souligner [1] qu'il n'est pas évident que le bump soit lié à un processus classique d'extinction, que ce soit de l'absorption pure ou de l'extinction par un type de particules particulier. Le bump me semble être une interruption de la lumière diffusée seule, et ne pas affecter la lumière directe de l'étoile. Cela est illustré par la paramétrisation de HD46223 (figure 1) et par la courbe de Seaton (figure 1 de [1]). Dans les deux cas il est curieux de constater que le prolongement de l'extinction visible dans l'UV, passe près de l'extremum du bump. Cela peut être une coïncidence, mais elle est suffisamment troublante pour ne pas s'interroger. Ici encore, un projet engageant des observations d'un grand nombre d'étoiles derrière un même nuage serait utile pour comprendre comment le bump se crée. Il est difficile dans l'état actuel des choses d'avoir une idée déterminée sur cette question.

# 7. Conclusion

L'interprétation standard de la courbe d'extinction s'est établie à partir d'une lecture phénoménologique immédiate de ses différentes parties. Les efforts développés pour donner un sens physique à cette interprétation ont conduit au développement de modèles de poussières sophistiqués mais jamais satisfaisants. Dans une précédente note, je me suis appuyé sur des raisonnements de bon sens pour montrer que cette lecture trop rapide n'est pas en accord avec l'observation. Je montre ici que le pendant de l'interprétation standard, à savoir que la lumière qui nous parvient d'une étoile rougie est contaminée par de la lumière diffusée, résout de façon simple les principaux problèmes posés par la courbe d'extinction: son interprétation, le sens physique de sa paramètrisation mathématique.

La lumière qui nous parvient d'une étoile rougie est contaminée par de la lumière diffusée. Cette contamination est importante parce que la diffusion se fait de façon cohérente. Vu de l'observateur l'angle à l'intérieur duquel la diffusion cohérente s'effectue est très faible, de l'ordre de $10^{-8}$" de l'étoile observée. La lumière diffusée domine largement la lumière directe de l'étoile dans l'UV, si le rougissement est suffisant: dans ce cas, dans l'UV lointain, on observe la lumière diffusée et non la lumière directe de l'étoile.

La courbe d'extinction réelle de la lumière par la matière interstellaire est une ligne droite du proche infra-rouge jusque dans l'UV lointain.

La comparaison de la loi UV de diffusion dans les nébuleuses et de celle trouvée pour la lumière diffusée qui contamine le spectre des étoiles rougies montre que deux types de particules sont nécessaires pour expliquer l'extinction de la lumière par la matière interstellaire. Des gros grains, qui diffusent vers l'avant avec une section efficace en $1/\lambda$, de petites particules (gaz ou poussières), qui diffusent selon la loi de Rayleigh en $1/\lambda^4$. Dans ce contexte, contrairement aux idées reçues, il n'y a plus lieu de supposer des variations des propriétés moyennes de la poussière interstellaire suivant le nuage considéré.

Trois problèmes n'ont pu être étudié de façon précise. Il s'agit d'abord du bump à 2200 Å. Je doute, mais le débat reste ouvert, qu'il s'agisse, comme il est communément admis, d'un processus d'extinction par un type spécifique de particules. Il semble que seule la lumière diffusée soit interrompue par le bump, qui dans ce cas n'affecte pas la lumière directe. Ce travail montre également la possibilité d'obtenir une représentation mathématique des courbes d'extinction en accord avec leur interprétation physique, et de diminuer le nombre de paramètres nécessaires à leur ajustement, qui sont aujourd'hui au nombre de six dans la décomposition de Fitzpatrick et Massa. Enfin la nature des petites particules qui diffusent de façon cohérente la lumière d'une étoile rougie reste à déterminer. Toutefois, l'hydrogène apparait comme une solution naturelle et en excellent accord avec les observations.